\documentstyle[twocolumn,pra,aps,psfig,graphicx]{revtex}
\begin{document}
%\twocolumn[\hsize\textwidth\columnwidth\hsize\csname@twocolumnfalse\endcsname
\draft

%\noindent
%{\bf Title:} (for cond-mat submission)

%\noindent 
%The critical exponent of the localization length at the Anderson
%transition in 3D disordered systems is larger than $1$

%\noindent
%{\bf Abstract:} (for cond-mat submission)

%In a recent communication to the cond-mat archives, Suslov
%[cond-mat/0105325] severely criticizes a multitude of numerical
%results obtained by various groups for the critical exponent $\nu$ of
%the localization length at the disorder-induced metal-insulator
%transition in the three-dimensional Anderson model of localization as
%``entirely absurd'' and ``evident desinformation''.  These claims are
%based on the observation that there still is a large disagreement
%between analytical, numerical and experimental results for the
%critical exponent. The author proposes, based on a ``simple procedure
%to deal with corrections to scaling'', that the numerical data support
%$\nu\approx 1$, whereas recent numerical papers find $\nu = 1.58 \pm
%0.06$.

%As we show here, these claims are entirely wrong. The proposed scheme
%does neither yield any improved accuracy when compared to the existing
%finite-size scaling methods, nor does it give $\nu\approx 1$ when
%applied to high-precision data. Rather, high-precision numerics with
%error $\varepsilon\approx 0.1\%$ together with all available
%finite-size-scaling methods evidently produce a critical exponent
%$\nu\approx 1.58$.

%\clearpage

%%%%%%%%%%%%%%%%%%%%%%%%%%%%%%%%%%%%%%%%%%%%%%%%%%%%%%%%%%%%%%%%%%%%%%%%
%
% Introduction
%
%%%%%%%%%%%%%%%%%%%%%%%%%%%%%%%%%%%%%%%%%%%%%%%%%%%%%%%%%%%%%%%%%%%%%%%%

\noindent
In a recent communication to the cond-mat archives, Suslov
\cite{Sus01} severely criticizes a multitude of numerical results
obtained by various groups for the critical exponent $\nu$ of the
localization length at the disorder-induced metal-insulator transition
(MIT) in the three-dimensional (3D) Anderson model (AM) of
localization as ``entirely absurd'' and ``evident desinformation''.
These claims are based on the observation that there still is a large
disagreement between analytical, numerical and experimental results
for the critical exponent \cite{KraM93}. The author proposes, based on
a ``simple procedure to deal with corrections to scaling'', that the
numerical data support $\nu\approx 1$, whereas recent numerical
papers find $\nu = 1.58 \pm 0.06$ \cite{SleO99a,SleO99b,OhtSK99}.

As we show below, these claims are entirely wrong. The proposed scheme
does neither yield any improved accuracy when compared to the existing
finite-size scaling (FSS) methods, nor does it give $\nu\approx 1$
when applied to high-precision data.

FSS at the Anderson MIT has a noteworthy history, reaching a first
peak with the seminal papers of Pichard/Sarma \cite{PicS81a,PicS81b}
and MacKinnon/Kramer \cite{MacK81,MacK83}.  Especially in Ref.\ 
\cite{MacK83}, the groundwork for a reliable, numerical FSS procedure
was laid and scaling curves could be constructed that proved the
existence of an MIT in 3D. In these and later studies based on the
same analysis technique \cite{KraM93}, the critical exponent $\nu$, as
estimated from the divergence of the infinite-size localization and
correlation lengths $\xi(W)$ at the transition $W_c$, i.e., $\xi
\propto |1-W/W_c|^{-\nu}$, is systematically {\em underestimated},
since the divergent nature at the transition can only be poorly
captured by FSS of data obtained for small system sizes and large
errors $\varepsilon$ in these finite-size data.  However, as more
powerful computers became availably in the last decade, one observed a
trend towards larger values of $\nu \approx 1.35$
\cite{KraS96,SchKM89,KraBMS90,HofS93b} for $\varepsilon \leq 1\%$.

\begin{figure}[t]  
\includegraphics[width=0.9\columnwidth]{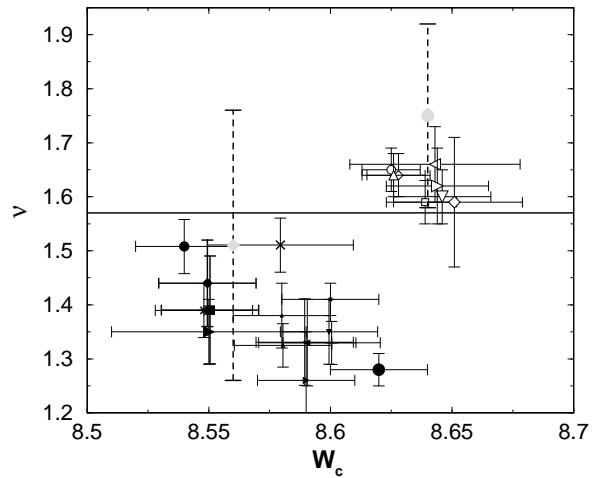}
\caption{\label{fig-nu-wc}       
  Results for $W_c$ and $\nu$, for the {\em anisotropic} AM
  \protect\cite{MilRSU00,MilRS99a} using SOS of TM data (open symbols)
  and ELS data (filled symbols) for various fit functions. The error
  bars show the $95\%$ confidence intervals.  The accuracy of TM
  localization lengths data is an order of magnitude higher than that of 
  the ELS data and the system sizes of TM data are larger than for ELS
  data, giving systematically larger $\nu$ values for the former. The
  goodness of a fit is reflected in the size of the symbol. The $2$
  thick error bars mark high quality ELS fits for large system sizes.
  The gray $\circ$ and $\Box$ and the corresponding error bars (dashed
  lines) represent $\nu_{\text{Suslov}}$ of TM data and ELS data for
  the anisotropic AM, respectively. The solid line marks the result of
  \protect\cite{SleO99a}.  }
\end{figure}

In 1994, high-precision data ($\varepsilon \leq 0.2\%$) showed a hitherto
neglected systematic shift of the transition point $W_c$ with
increasing system size.  Taking this into account phenomenologically,
$\nu = 1.54\pm0.08$ was found \cite{Mac94}. A subsequent approach by
Slevin/Ohtsuki \cite{SleO99a,SleO99b,OhtSK99} incorporated these
shifts as irrelevant scaling variables and further allowed for
corrections to scaling due to nonlinearities. With higher-precision
data ($\varepsilon\approx 0.1\%$), they found $\nu=1.57\pm 0.04$. Further
results for, e.g., the AM with anisotropic hopping
\cite{MilRSU00,MilRS99a,MilRS01}, the off-diagonal AM
\cite{CaiRS99,BisCRS00}, the AM in a magnetic field
\cite{ZhaK98,ZhaK97}, confirmed this value of $\nu$ within the error
bars (see Fig.\ \ref{fig-nu-wc}). Also, $\nu$ is identical for the MIT
as a function of disorder or energy \cite{CaiRS99,BisCRS00}. We
emphasize that a properly performed Slevin/Ohtsuki scaling (SOS)
procedure needs to assume various fit functions and that the final
estimates are to be suitably extracted from many such functional forms
\cite{MilRS99a,MilRS01,CaiRS99}; bootstrap
\cite{SleO99a,SleO99b,OhtSK99} or Monte Carlo methods
\cite{MilRS99a,MilRS01,CaiRS99} then need to be employed for a precise
estimate of error bars.

We have tested the method proposed by Suslov \cite{Sus01} first with the
transfer-matrix (TM) data of Refs.\ \cite{MilRSU00,CaiRS99,BisCRS00}
with $\varepsilon\leq 0.1\%$; we find $\nu_{\text{Suslov}}= 1.75\pm 0.17$
for the anisotropic and $1.55\pm 0.04$ for the random-hopping AM.  The
SOS gives $\nu=1.61 \pm 0.07$ \cite{MilRSU00} and $\nu=1.54\pm0.03$
\cite{CaiRS99,BisCRS00}, respectively.
Using for a second test energy-level-statistics (ELS) data
\cite{MilRS99a} with $\varepsilon\approx 1\%$, we find
$\nu_{\text{Suslov}}=1.51 \pm 0.25$, whereas SOS gives $1.45\pm0.2$
\cite{MilRS99a}.
Last, for artificially generated data with precisely known $W_c=16.5$
and varying $\nu\in [0.5,2.0]$ the results of the Suslov method are
comparable to the results of the MacKinnon/Kramer FSS and slightly less
reliable than the SOS. 
%%%%%%%%%%%%%%%%%%%%%%%%%%%%%%% FIGURE %%%%%%%%%%%%%%%%%%%%%%%%%%%%%%%%%%%%%%
%%%%%%%%%%%%%%%%%%%%%%%%%%%%%%% END OF FIGURE %%%%%%%%%%%%%%%%%%%%%%%%%%%%%%%

We conclude that the method proposed by Suslov also yields $\nu
\approx 1.58$ and not $\nu\approx 1$ for the MIT of the AM.

In principle, the Suslov method does not need to assume any functional
form of the FSS curves just as the MacKinnon/Kramer method. As a
numerical tool, the Suslov method is not unreasonable, but certainly
not better than the established methods: it does not take into account
the systematic shift due to irrelevant scaling variables, it relies on
an a-priori knowledge of $W_c$ and inherently produces rather large
error bars for the critical exponent. We note that Suslov in his
numerical test \cite{Sus01} used data for $3$ small system sizes
$6^3$, $12^3$ and $28^3$, while currently sizes $\sim 50^3$ (for ELS)
and $18^2\times 10^8$ (for TM) are standard. It is evident to people
with experience in FSS that Suslov's erroneously small $\nu$ is due
to his use of too few and too small system sizes. 

In conclusion, high-precision numerics with error $\varepsilon\approx
0.1\%$ together with all the above mentioned FSS methods produce a
critical exponent $\nu\approx 1.58 > 1$ for 3D. The numerical values
of $\nu$ for dimensions $2 < d < 3$ \cite{SchG96} and $4$
\cite{SchG96,ZhaK98} remain valid, they are certainly not ``entirely
absurd'' although there is only limited agreement with the field
theoretic approach \cite{KraM93}.
Similarly precise data are much harder to obtain for our experimental
colleagues, but recent advances in this direction show a clear trend
towards increasing $\nu$ \cite{WafPL99,ItoWOH99}.

\vspace{5pt}
\noindent
P.\ Cain, M.L.\ Ndawana, R.A.\ R\"{o}mer and M.\ Schreiber\\
Institut f\"{u}r Physik, Technische Universit\"{a}t\\
D-09107 Chemnitz, Germany

\vspace{5pt}
\noindent
($Revision: 1.12 $, printed: \today)

\vspace{-10pt}
%\bibliographystyle{prsty}
%\bibliography{bibliograph}

\end{document}